\documentclass[a4paper,fleqn, usenatbib]{mnras}

\usepackage{newtxtext,newtxmath}

\usepackage[T1]{fontenc}
\usepackage{ae,aecompl}

\usepackage{graphicx}	
\usepackage{amsmath}	
\usepackage{amssymb}	

\title[Extragalactic Satellite Bands]{Effects of coplanar satellite bands on galactic disc evolution}

\author[A. Criswell and C. Struck]{
Alexander Criswell,$^{1}$\thanks{E-mail: criswell@alumni.iastate.edu (AC); curt@iastate.edu (CS)} Curtis Struck$^{1}$\\
$^{1}$Dept. of Physics and Astronomy, Iowa State Univ., Ames, IA 50011 USA}

\date{Accepted XXX. Received YYY; in original form ZZZ}

\pubyear{2018}

\begin{document}
\label{firstpage}
\pagerange{\pageref{firstpage}--\pageref{lastpage}}
\maketitle

\begin{abstract}
Small dwarf companions have been long thought to have minimal influence on their host galaxy's evolution without undergoing direct impacts to the host's disc. However, in light of recent discoveries of coplanar, corotating satellite structures around the Milky Way, Andromeda, and Centaurus A, we use an N-body/test particle simulation to show that low-mass dwarf satellites within such structures are able to exert significant influence on their host's disc, driving spiral waves and inducing stellar scattering. This is accomplished through quasi-periodic alignments of multiple small satellites within the structure that emulate the gravitational influence of a single, larger satellite such as Sagittarius Dwarf or the Large Magellanic Cloud. We find that the coplanar, corotational nature of such structures allows for repeated alignments on short enough timescales to overcome damping within the disc, and in a consistent enough fashion to continually drive spiral waves over the course of 2 Gyr of simulation time. The spirals driven by this phenomenon tend to be flocculent and many-armed due to the irregular intervals over which alignments occur. We additionally find that while the aligned satellites are able to induce noticeable thickening of the disc, their ability to drive surface density profile evolution is secondary to other effects.
\end{abstract}




\section{Introduction}

The interaction between satellite dwarf galaxies and their host galaxy is a topic of much recent interest. Research into the effects these satellites have on host galaxy evolution can be viewed as falling into two main categories. The first branch focuses on the effects of a single, high-mass satellite similar to to the Sagittarius dwarf galaxy (Sgr) or the Large Magellanic Cloud (LMC). \citet{Sellwood1998} were able to show that an infalling satellite with a mass approximately 5\% of that of its host galaxy could excite ''disc-bending waves'', causing vertical thickening of the galactic disc. \citet{Bailin2004} found that an LMC-sized satellite dwarf was able to cause disc warps in their model Milky Way. Shortly thereafter, \citet{Weinberg2004} showed that a single, large, infalling satellite could interact with Lindblad Resonances within the disc, but changes derived from these resonances tended to be transient, while lasting effects stemmed from the angular momentum transfer involved in the satellite-galaxy collision. Later, \citet{Quillen2009} placed a $6x10^{9}M_{\sun}$ satellite in an eccentric orbit with a pericentre <10kpc from the galactic centre and thus ensured regular disc impacts. These close flybys and regular impacts resulted in disc warps and induced flocculent spiral waves. \citet{Purcell2011} modeled the effects produced in the Milky Way by the infall of Sgr, yielding spiral wave formation, enhanced bar evolution, and flaring at the edges of the Milky Way's disc. \citet{Gomez2012} and \citet{D'Onghia2016} also simulated the Sgr impact, producing vertical density waves and disc thickening, respectively. \citet{Widrow2015} used a similar large satellite to that of \citet{Quillen2009}, and in addition to observing the same spirals and thickening as their predecessors, were able to show that the effects of a single perturbance are quickly damped away within the disc and repeated, cyclic perturbances are necessary to induce long-term effects.
 
The second avenue of research looks at the cumulative effects of small satellites such as the average dwarf galaxy. \citet{Quinn1993} established that satellite mergers lead to heating of the host galaxy’s disc; \citet{Bird2011} further explored the effects of direct satellite bombardment on radial migration, treating each impact as a perturbation, and showing that such bombardment can cause stellar scattering. \citet{D'Onghia2016} also investigated the case of \textasciitilde1000 $10^{7}M_{\sun}$ satellites bombarding the disc as well as that of \textasciitilde100 $2x10^{8}-2x10^{10}M_{\sun}$ isotropically distributed dark subhaloes in orbits of varying eccentricity. In the former case, they found that the low-mass satellite bombardment produced no major effects. In the latter case, however, they observed disc heating and flaring. An important demonstration of the effects of small, dark matter satellites came from \citet{Dubinski2008}, who found that direct impacts of satellites in the $2x10^{8}-2x10^{10}M_{\sun}$ range can induce spiral waves and bars in the disc. 
Conversely, little work has been focused on the effects of non-impacting small satellites - understandably so, as individual low-mass satellites have generally been considered unable to produce noticeable, long-term effects without direct disc impacts. They are simply too small and fail to provide enough of a perturbation to significantly influence the disc. However, the recent discoveries of great planar corotating satellite alignments - phenomena we call `band structures' - around the Milky Way \citep{Kroupa2005,Pawlowski2012}, Andromeda \citep{Koch2006,Ibata2013}, and Centaurus A \citep{Mueller2018} shed a new light on how small satellites may have a larger impact on the evolution of their host galaxy than their size might suggest. In \citet{Kroupa2005, Koch2006, Mueller2018}, we see that these band structures consist of between 14 and 24 dwarf satellites, with individual satellite orbital distances from their respective hosts ranging from 3 kpc to 900 kpc, with most lying in the 20-100 kpc range. Additionally, obervations of the structure around Andromada \citep{Collins2015} have shown that dwarf galaxies in these structures have masses on the order of $10^8$ to $10^{10}\  M_{\sun}$, with most falling in the $10^{9}\ M_{\sun}$ range.\\
We show that due to their corotation and different orbital periods, the component satellites of these band structures experience periodic chance alignments, resonantly perturbing the host galaxy disc. Thus, the satellites in the band structure are able to induce phenomena in the disc previously thought to be the realm of high-mass satellites or direct satellite impacts, such as flocculent or multi-armed spiral waves, disc thickening, and disc profile evolution.

\section{Methods}

In order to investigate this phenomenon, we utilized a heavily modified version of an original MATLAB \citep{MathWorks2018} test particle code used in \citet{Struck2017}. The code was originally designed to use fixed-orbit `clumps' within the disc to explore the effects of such objects on galactic profile evolution. Generally, we create a disc of \textasciitilde17,000 non-self-gravitating `star' particles in a combined disc/halo potential. The mechanics of the disc/halo potential are unchanged from the original code. In its current iteration, the code has been retooled to model satellite dwarf galaxies that are allowed to orbit freely about the host galaxy. Gravitational interactions within the simulation are handled by two main integrator functions, both of which are driven by the MATLAB ode23 routine, a four-stage, explicit, 2nd/3rd order Runge-Kutta ordinary differential equation solver. The first function governs the disc/halo potential, whereas the second oversees the gravitational effects of the band structure satellites on particles in the disc. The band structure satellites orbit in the disc/halo potential, but do not interact with each other.

Initially, the satellite dwarfs were placed into a band around the central disc; the average band radius and average band angular widths were variable, but the initial radial and azimuthal band widths remained constant at 5 kpc and 22.5 degrees, respectively. At the beginning of each run, 20 satellites were randomly distributed within these width constraints and given initial velocities designed to set them in low-eccentricity elliptical orbits corresponding to their initial positions. Over the course of our work, we varied the masses and orbital radii of the band structure satellites; the relevant values for the runs performed in this study can be found in Table~\ref{tab:runvartable}. With these initial conditions, the simulation subjected the band structure satellites to the influence of the combined satellite/halo potential throughout the run, allowing the dwarfs to orbit freely about the host. This naturally allows for precession of the satellite orbits and ensures that any periodic alignments are not merely a result of constrained satellite orbits.

While some observed band structures include slightly more than our 20 satellites, these structures often contain some satellites with orbital radii much too great to be relevant to this study. To account for this, we chose a reduced satellite count, simulating only the satellites reasonably close to the disc. Initially, we used larger satellite masses and smaller band radii than observed in an effort to study the effects in a short time. Over the course of the study, we introduced lower-mass satellites and larger band radii, bringing the simulation in-line with observations \citep{Collins2015,Kroupa2005,Pawlowski2012,Koch2006,Ibata2013,Mueller2018}. The later, more realistic runs produced results that while predictably lessened in amplitude, were consistent with those seen in earlier, larger-mass/smaller-radius runs. The low-mass satellites are similar to those discussed in \citet{Collins2015}, adjusted to account for the original values being given at half-light radius. The high-mass satellites are similar to Andromeda's NGC 147 and NGC 185 or the Large and Small Magellanic Clouds. Many of the runs were performed with a band tilt angle of zero degrees, simulating a polar orbital plane; this is largely motivated by the polar nature of the band structures around the Milky Way and Andromeda. This polar configuration is expected to have a lessened effect on the disc compared to a coplanar configuration \citep{D'Onghia2010}.\\

\begin{table}
	\centering
	\caption{Band radius given in terms of the inner edge; band angle given in terms of the minimum angle. The band initially extends beyond these values by 5kpc and 22.5° in radius and azimuth, respectively; over the course of a run, the satellites quickly spread out of their tight initial band due to precession and varied eccentricity. However, the structure retains its roughly coplanar/corotational nature. We follow the convention of measuring the azimuth angle from the vertical. Runs with a Satellite Mass value of ``Mixed'' use a range of satellite masses to simulate a more realistic case; the band consists of primarily lower-mass satellites and a few higher-mass ones.}
	\label{tab:runvartable}
	\begin{tabular}{cccc} 
		\hline
		Run \# & Satellite Mass & Band Radius & Band Angle\\
		& $10^9 M_{\sun}$ & kpc & \\
		\hline
		1 & 27.6 & 10.0 & 45° \\
		2 & 27.6 & 10.0 & 0° \\
		3 & 27.6 & 5.0 & 45° \\
		4 & 27.6 & 5.0 & 60° \\
		5 & 27.6 & 15.0 & 0° \\
		6 & 6.9 & 15.0 & 0° \\
		7 & 6.9 & 15.0 & 45° \\
		8 & 55.3 & 15.0 & 0° \\
		9 & 27.6 & 15.0 & 90° \\
		10 & 55.3 & 15.0 & 90° \\
		11 & 13.8 & 30.0 & 0° \\
		12 & 13.8 & 30.0 & 0° \\
		13 & 13.8 & 30.0 & 0° \\
		14 & 13.8 & 30.0 & 0° \\
		15 & 2.8 & 30.0 & 0° \\
		16 & 1.4 & 30.0 & 0° \\
		17 & 0.3 & 30.0 & 0° \\
		18 & Mixed & 30.0 & 0° \\
		19 & Mixed & 30.0 & 0° \\
		20 & 2.8 & 50.0 & 0° \\
		21 & Mixed & 50.0 & 0° \\
		22 & Mixed & 75.0 & 0° \\
		23 & Mixed & 100.0 & 0° \\
		24 & Mixed & 200.0 & 0° \\
		\hline
	\end{tabular}
\end{table}
The simulation outputs use units of convenience for time, distance, and mass - one Mass Unit is equivalent to $2.76 \times 10^8\ M_{\sun}$, one Distance Unit is equivalent to 0.5 kpc, and one Time Unit is equivalent to 9.8 Myr. Generally speaking, each instance of the simulation was run for 200 time units (\textasciitilde2 Gyr). During the preliminary work for this research, we found that pertinent effects would be observable within this timeframe, and mechanisms that failed to produce a noticeable impact on the galaxy during this time generally had a tertiary influence at best on host galaxy evolution. A few of the runs used in this article were ended earlier, but are included despite not reaching the 2 Gyr mark due to notable behavior that was observed even within their shortened lifespan.

\section{Results}
\subsection{Spiral Waves}
\begin{figure}
	\includegraphics[width=\columnwidth]{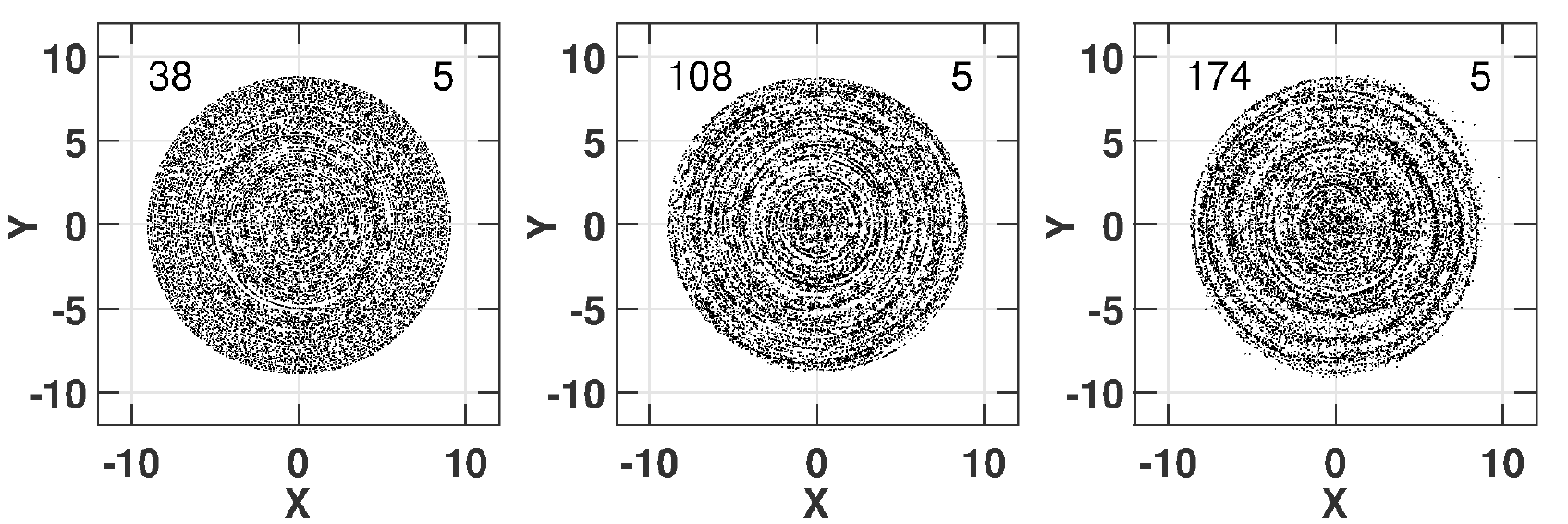}
    \caption{Spiral waves at 0.38 Gyr, 1.08 Gyr, and 1.74 Gyr for a run with $1.4x10^9 M_{\sun}$ satellites. (Run \#16) Numbers in the upper right of each panel give the satellite mass in code units.}
    \label{fig:figure1}
\end{figure}

\begin{figure}
	\includegraphics[width=\columnwidth]{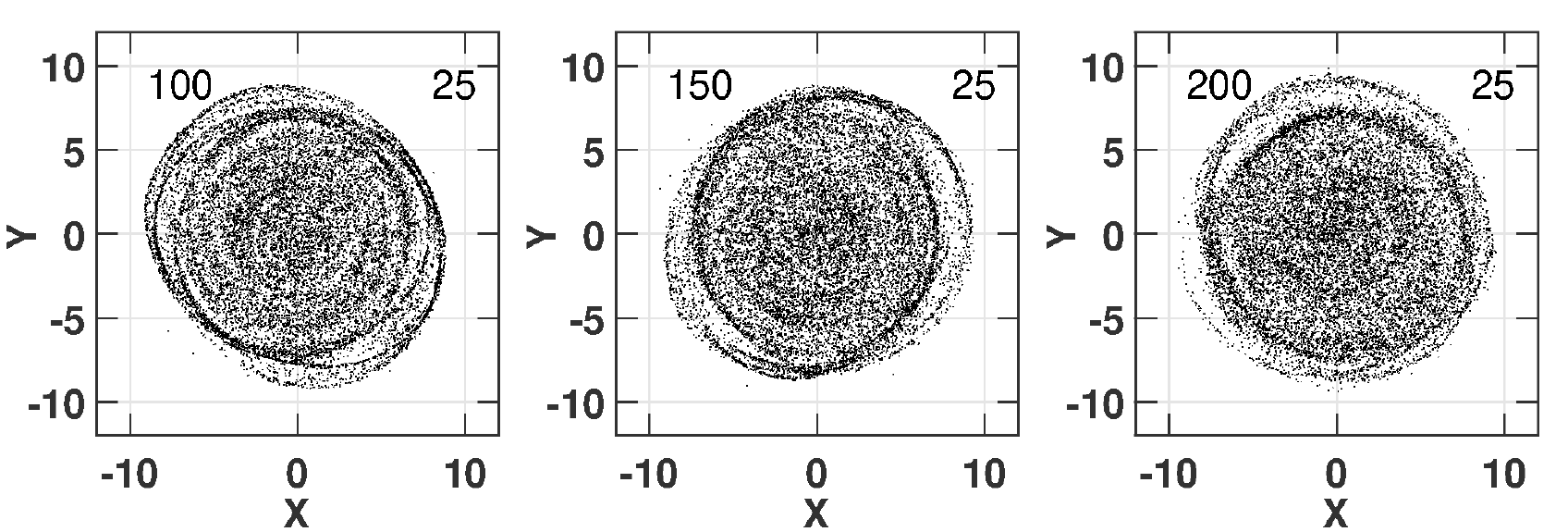}
    \caption{Spiral waves at 0.98 Gyr, 1.47 Gyr, and 1.96 Gyr for a run with $6.9x10^9 M_{\sun}$ satellites. (Run \#6)}
    \label{fig:figure2}
\end{figure}

\begin{figure}
	\includegraphics[width=\columnwidth]{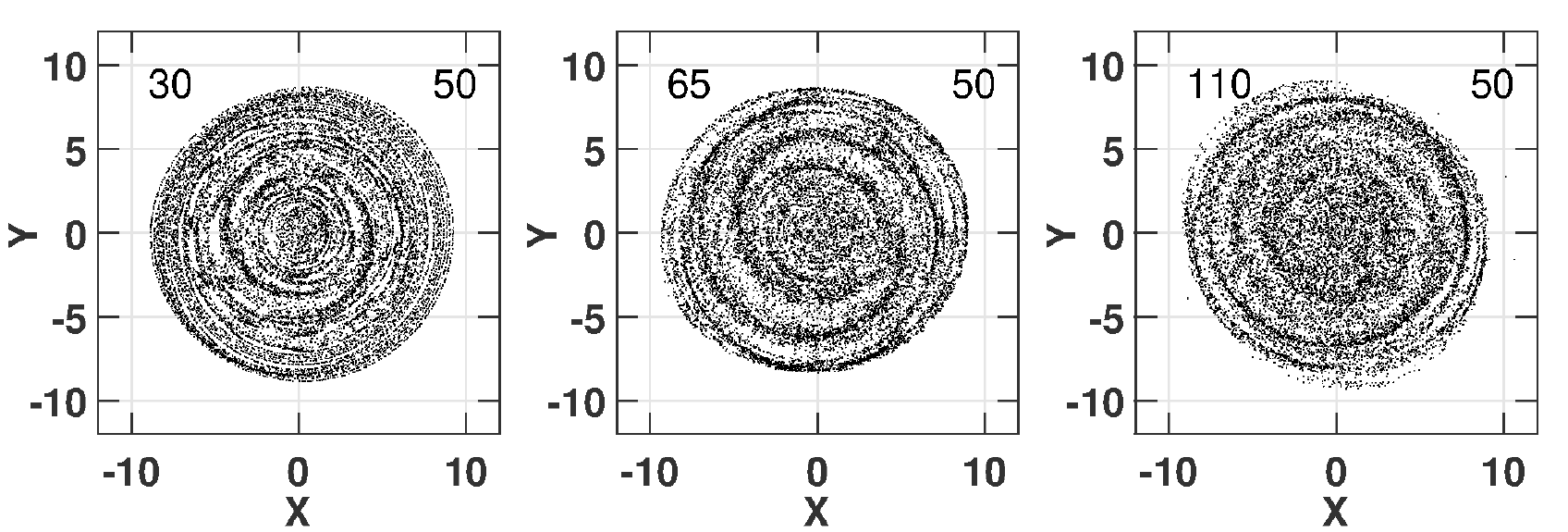}
    \caption{Spiral waves at 0.29 Gyr, 0.64 Gyr, and 1.08 Gyr for a run with $13.8x10^9 M_{\sun}$ satellites. (Run \#11)}
    \label{fig:figure3}
\end{figure}

In most of our runs the presence of a band structure consistently induces persistent spiral waves within the disc. The occasional spiral is not terribly notable, but the repeated occurrence and persistent nature of these spirals sets them apart from mere sporadic anomalies. Figs.~\ref{fig:figure1},~\ref{fig:figure2}, and~\ref{fig:figure3} each consist of 3 frames, showing this behavior at successive points in time for 3 different runs. 


\begin{figure}
	\includegraphics[width=\columnwidth]{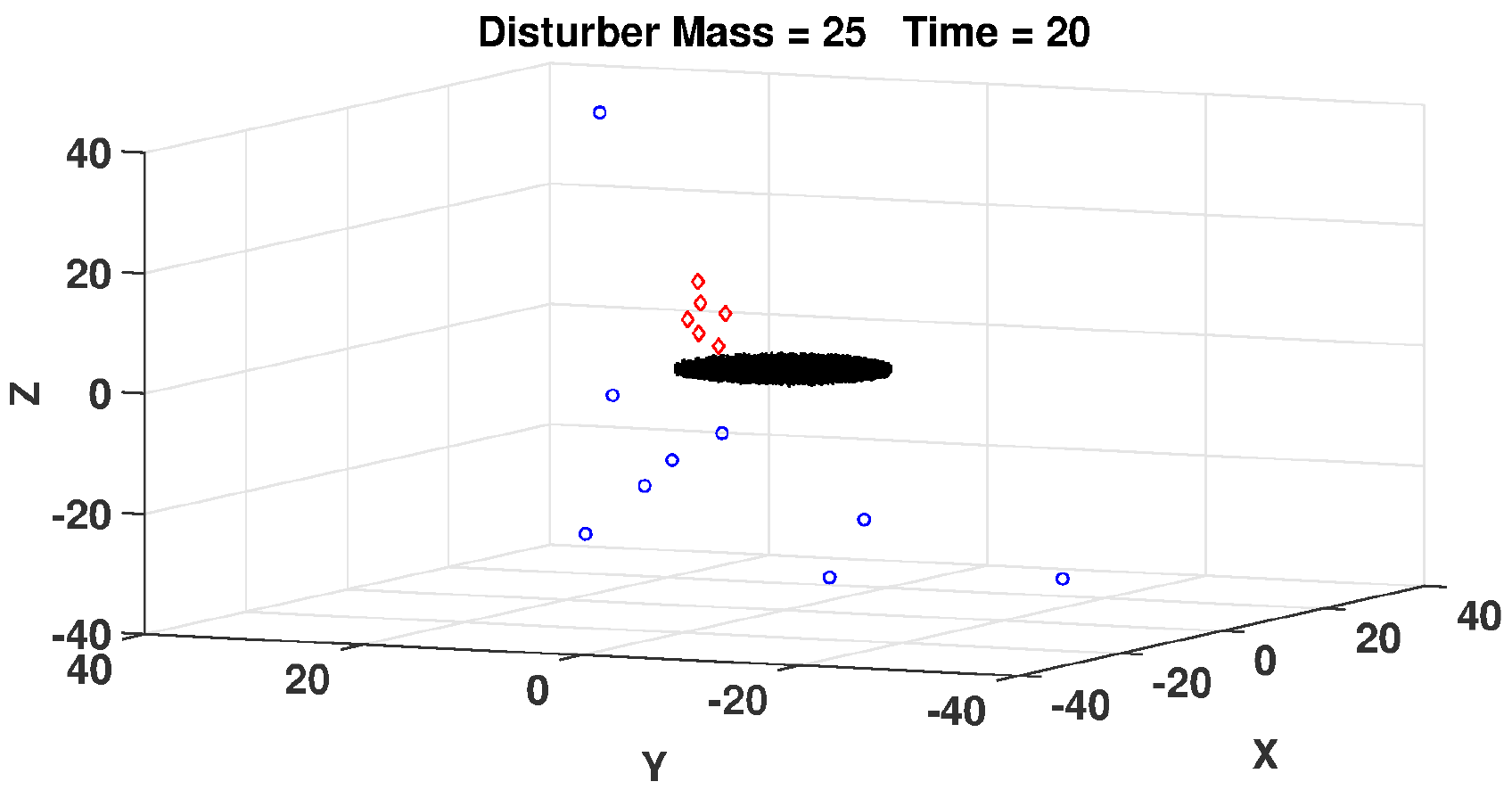}
    \caption{Visually identified satellite alignment at 196 Myr for a run with $6.9x10^9 M_{\sun}$ satellites. (Run \#6)}
    \label{fig:figure4}
\end{figure}

\begin{figure}
	\includegraphics[width=\columnwidth]{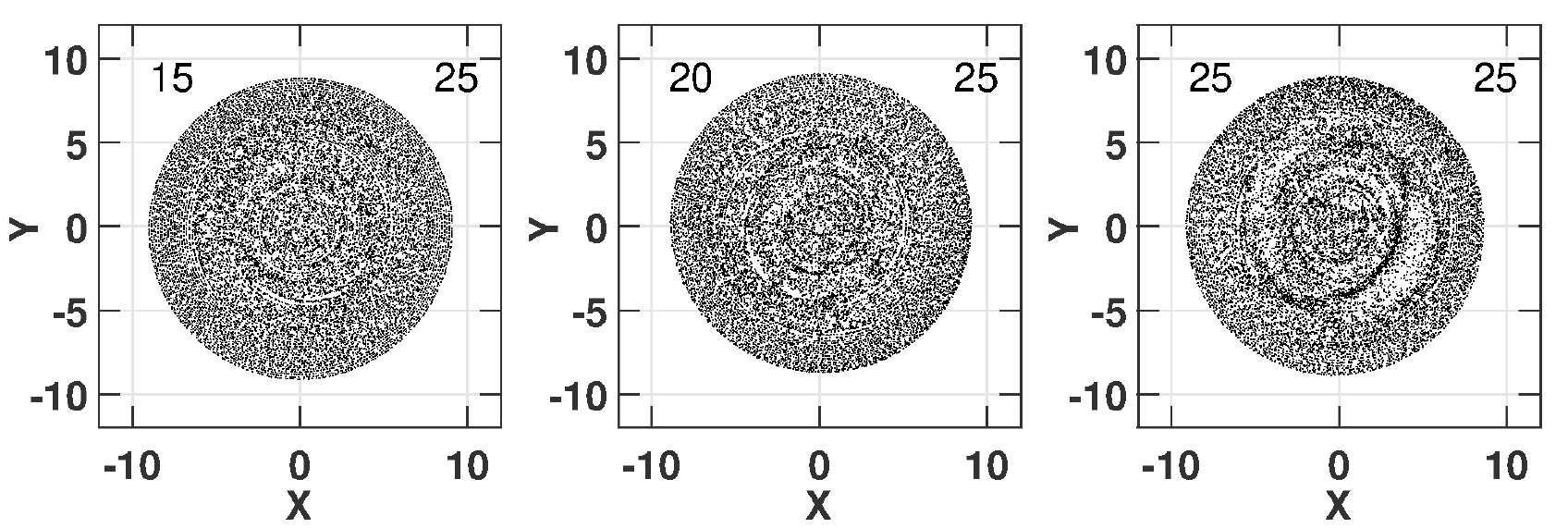}
    \caption{Spiral wave formation occuring over 98 Myr as a result of satellite alignment in Figure~\ref{fig:figure4}.}
    \label{fig:figure5}
\end{figure}

\begin{figure}
	\includegraphics[width=\columnwidth]{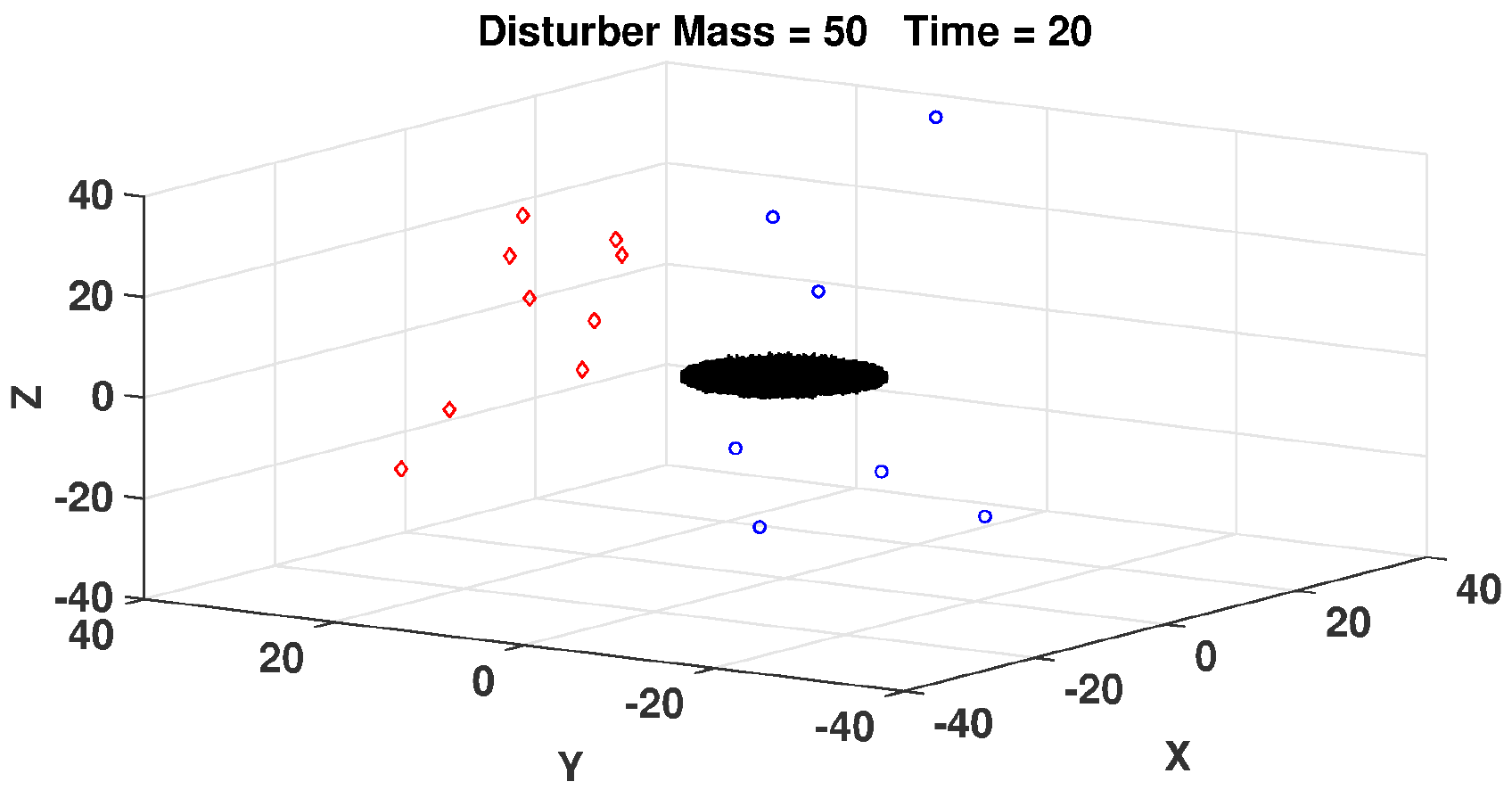}
    \caption{Visually identified satellite alignment at 196 Myr for a run with $13.8x10^9 M_{\sun}$ satellites. (Run \#11)}
    \label{fig:figure6}
\end{figure}

\begin{figure}
	\includegraphics[width=\columnwidth]{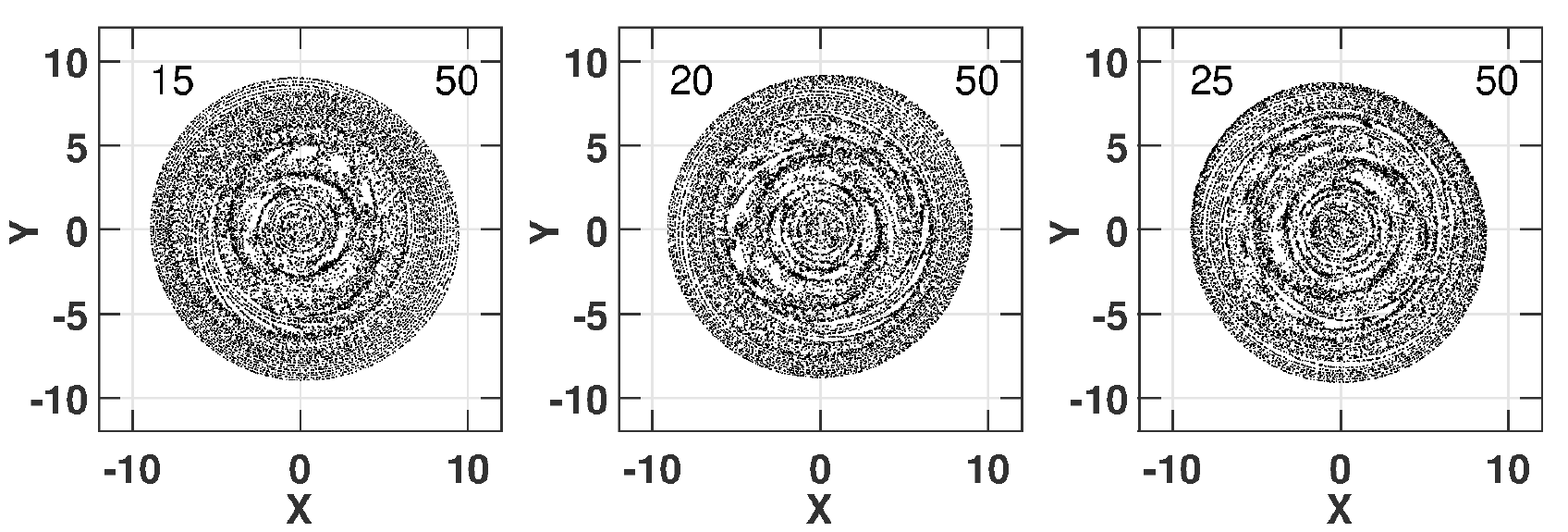}
    \caption{Spiral wave formation occuring over 98 Myr as a result of satellite alignment in Figure~\ref{fig:figure6}.}
    \label{fig:figure7}
\end{figure}

\begin{figure}
	\includegraphics[width=\columnwidth]{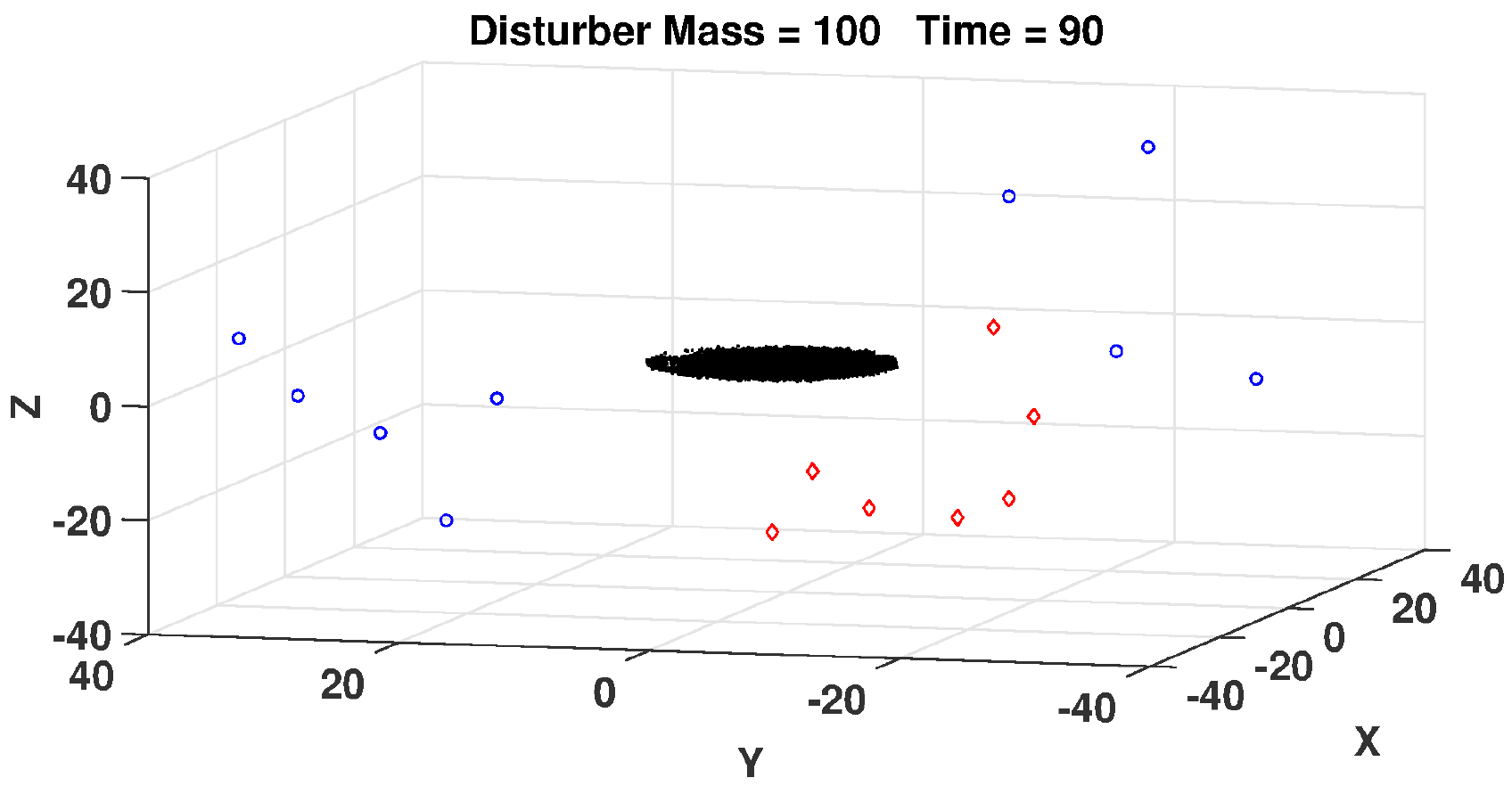}
    \caption{Visually identified satellite alignment at 882 Myr for a run with $27.6x10^9 M_{\sun}$ satellites.  (Run \#9)}
    \label{fig:figure8}
\end{figure}

\begin{figure}
	\includegraphics[width=\columnwidth]{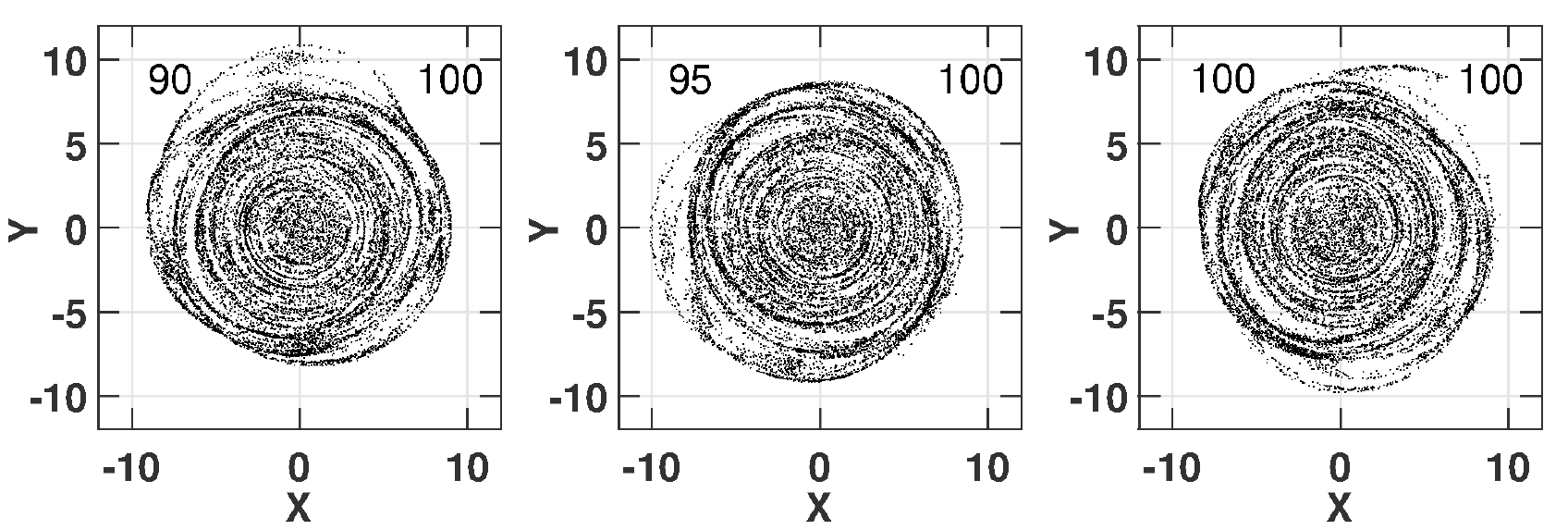}
    \caption{Spiral wave re-excitation occuring over 98 Myr as a result of satellite alignment in Figure~\ref{fig:figure8}.}
    \label{fig:figure9}
\end{figure}

The cause of these persistent spirals appears to be quasi-periodic alignments of satellites in the band structure. Due to their corotational nature and varying orbital periods, clusters of satellites will occasionally form in the band. These alignments have been consistently observed within the 10 Myr prior to a new spiral forming in the disc, as seen in three examples in Figs.~\ref{fig:figure4} through~\ref{fig:figure9}.


Previously, only large satellites in close orbits had been considered to have the capacity to drive disc spirals \citep{Quillen2009,Widrow2015}. If, as our models suggest, band structure alignments allow non-impacting small satellite galaxies to exhibit this behavior, then the combined gravitational attraction of the aligned satellites emulates that of a single, more massive satellite. We used this idea to create an alignment criterion. The criterion considers the combined gravitational force exerted by a number of satellites along a given positional vector and compares that value to same measure for a single satellite at a location given by the aforementioned position vector at a radius similar to that of \citet{Quillen2009} and a mass sufficient to induce spiral waves in the disc. If the former value is greater than the latter, the satellites are considered aligned. Specifically, this criterion can be written as,

\begin{equation}
     $$\sum_{i=1}^{n} \frac{Gm_i}{(proj_{\vec{R}}\vec{r_{i}})^{2}}\geq G\frac{M}{|\vec{R}|^2}$$
	\label{eq:aligncrit}
\end{equation}

\begin{figure}
	\includegraphics[width=\columnwidth]{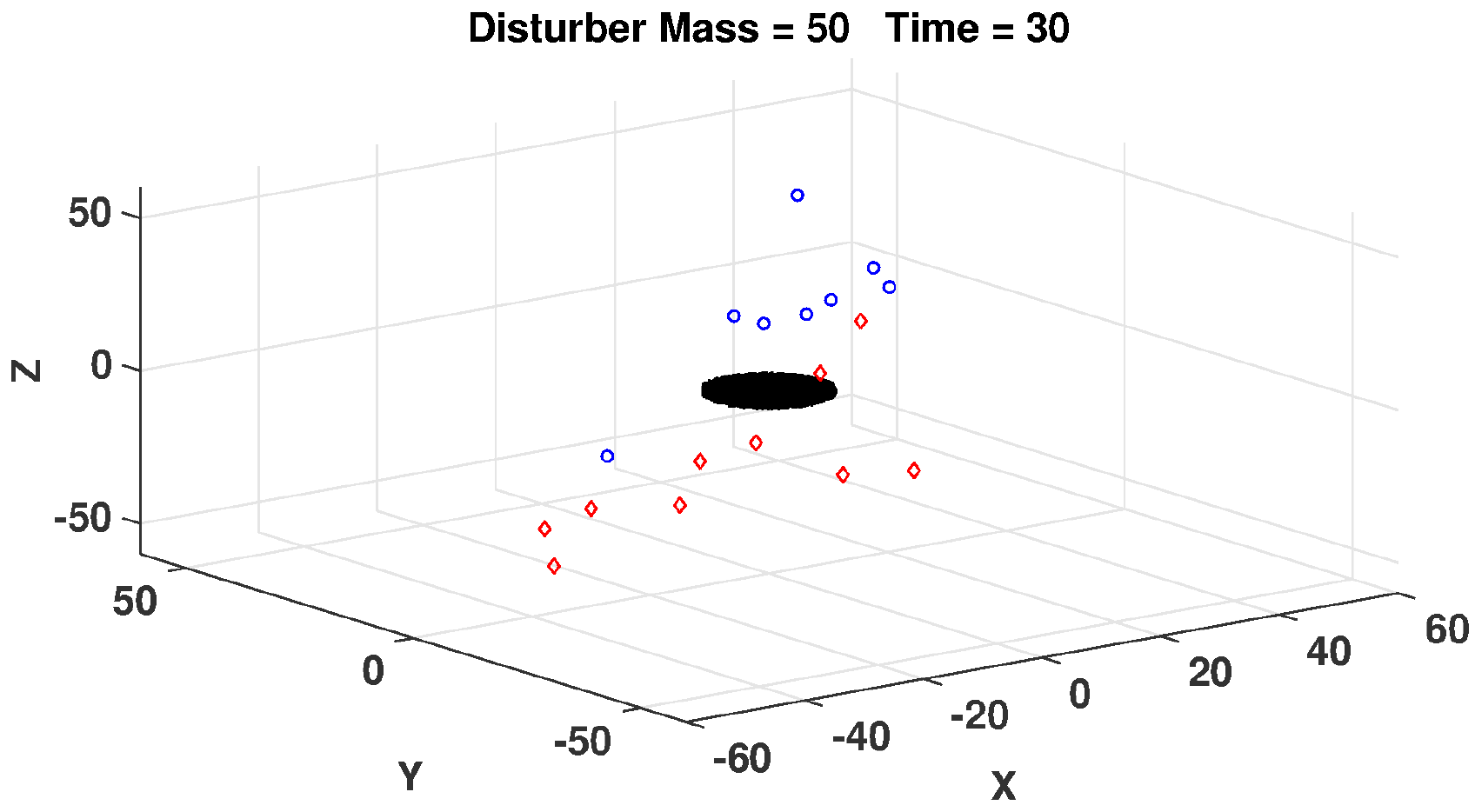}
    \caption{Satellite alignment as defined by alignment criterion at 294 Myr for a run with $13.8x10^9 M_{\sun}$ satellites. (Run \#13)}
    \label{fig:figure10}
\end{figure}

\begin{figure}
	\includegraphics[width=\columnwidth]{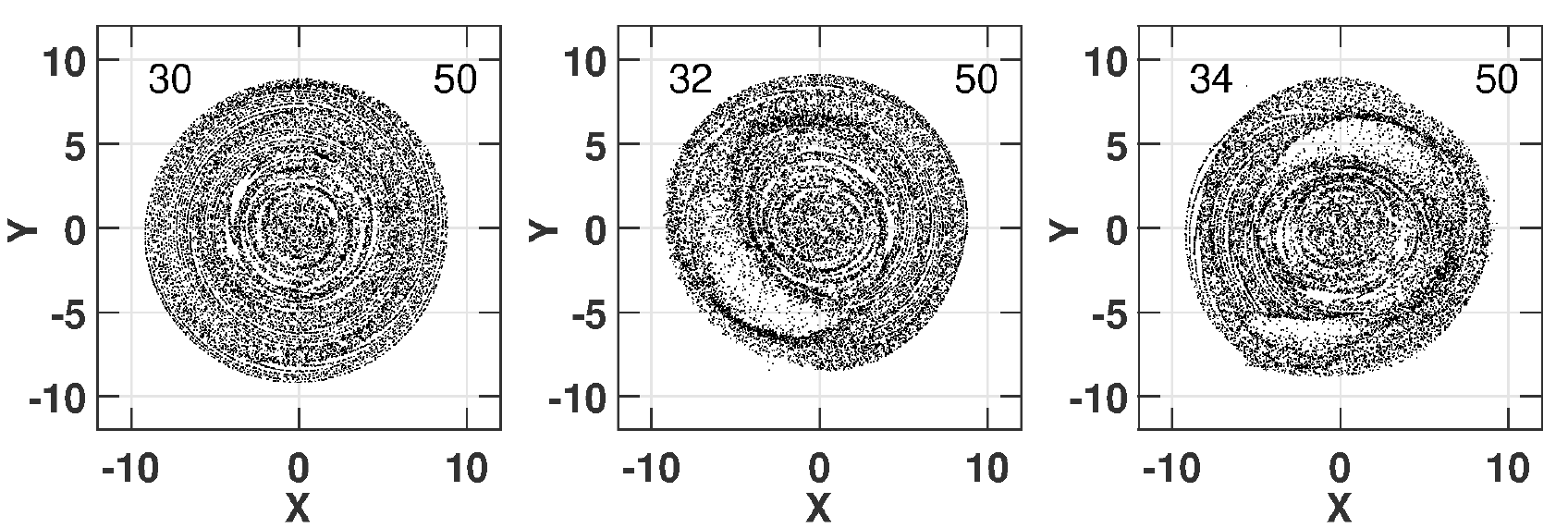}
    \caption{Spiral wave formation occuring over 39 Myr as a result of satellite alignment in Figure~\ref{fig:figure10}.}
    \label{fig:figure11}
\end{figure}

\begin{figure}
	\includegraphics[width=\columnwidth]{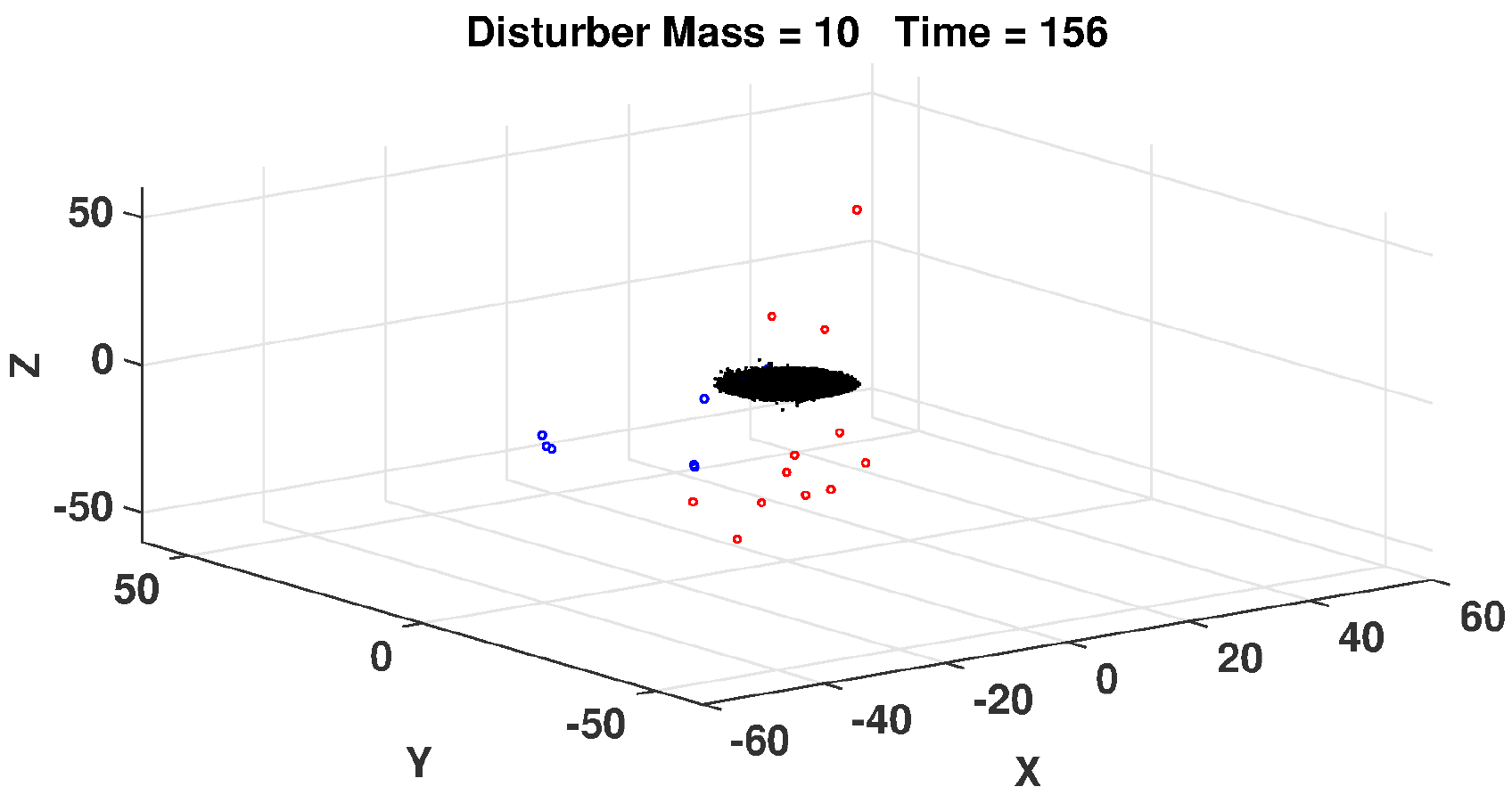}
    \caption{Satellite alignment as defined by alignment criterion at 1.53 Gyr for a run with $2.8x10^9 M_{\sun}$ satellites. (Run \#15)}
    \label{fig:figure12}
\end{figure}

\begin{figure}
	\includegraphics[width=\columnwidth]{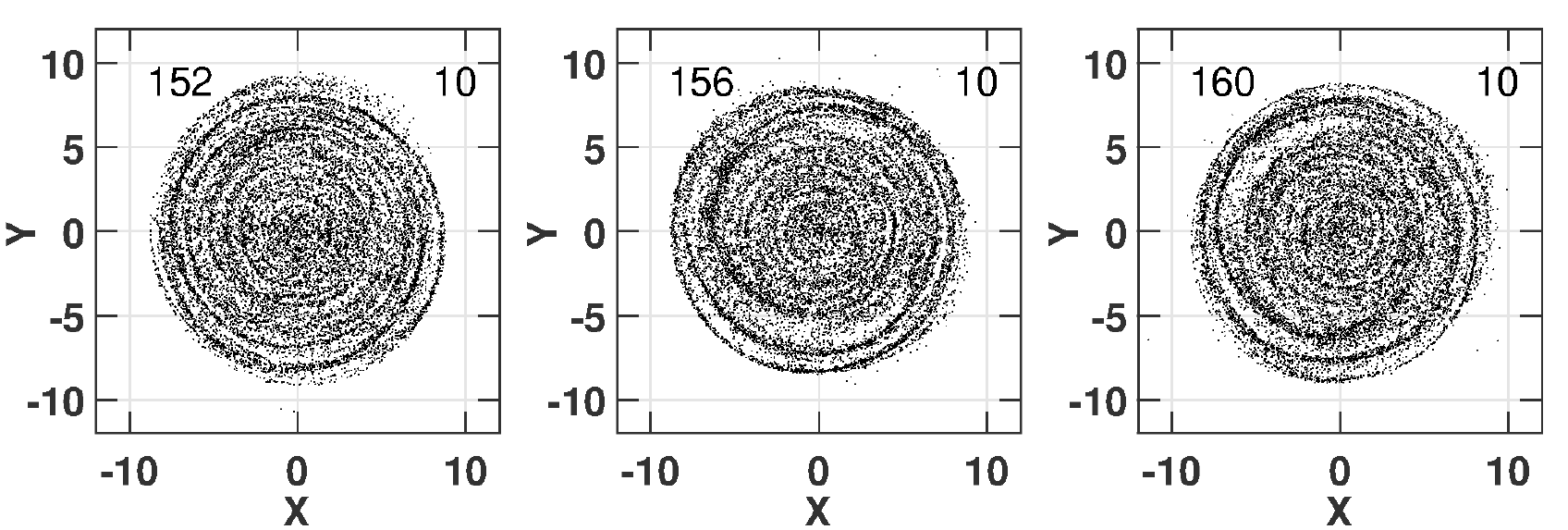}
    \caption{Spiral wave formation occuring over 78 Myr as a result of satellite alignment in Figure~\ref{fig:figure12}.}
    \label{fig:figure13}
\end{figure}

\begin{figure}
	\includegraphics[width=\columnwidth]{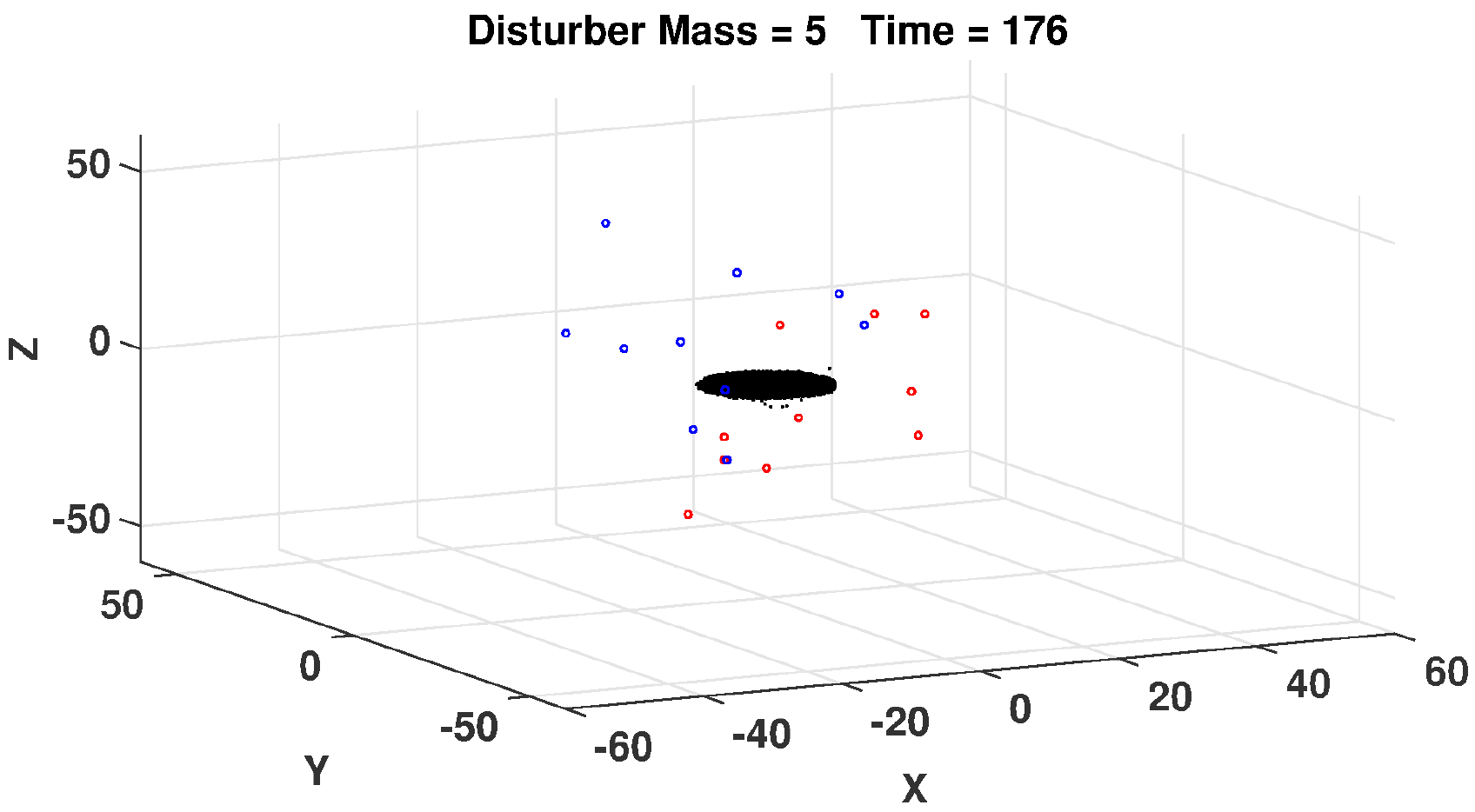}
    \caption{Satellite alignment as defined by alignment criterion at 1.72 Gyr for a run with $1.4x10^9 M_{\sun}$ satellites. (Run \#16)}
    \label{fig:figure14}
\end{figure}

\begin{figure}
	\includegraphics[width=\columnwidth]{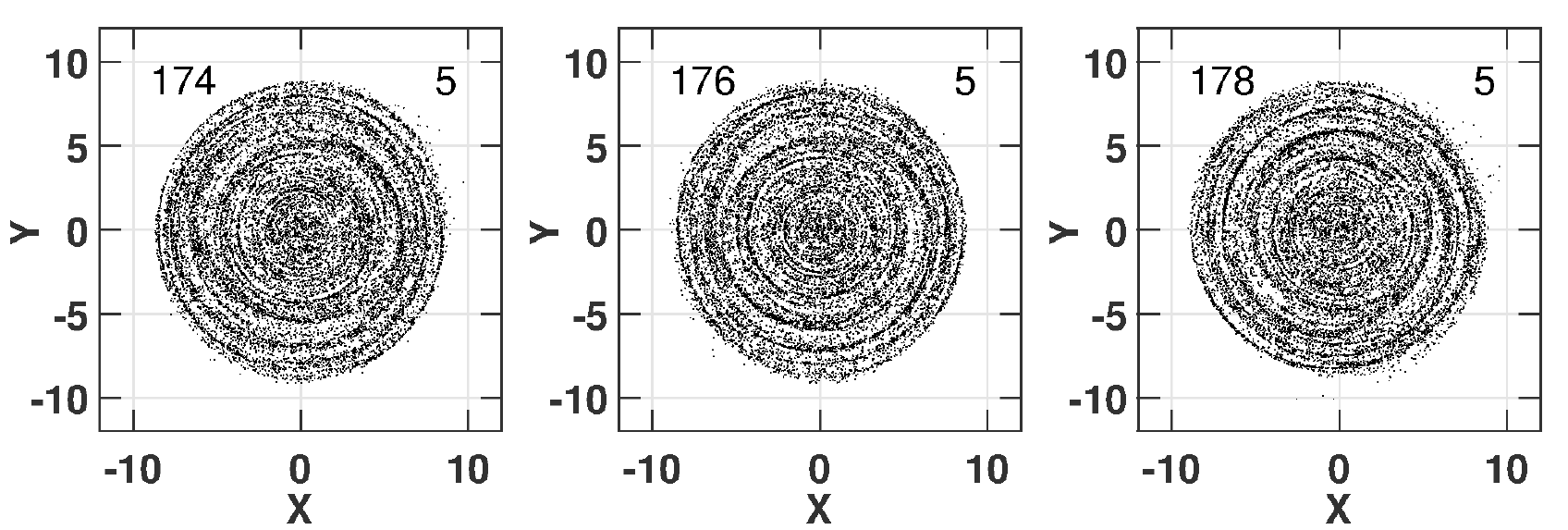}
    \caption{Spiral wave re-excitation occuring over 39 Myr as a result of satellite alignment in Figure~\ref{fig:figure14}.}
    \label{fig:figure15}
\end{figure}

\begin{figure}
\includegraphics[width=\columnwidth]{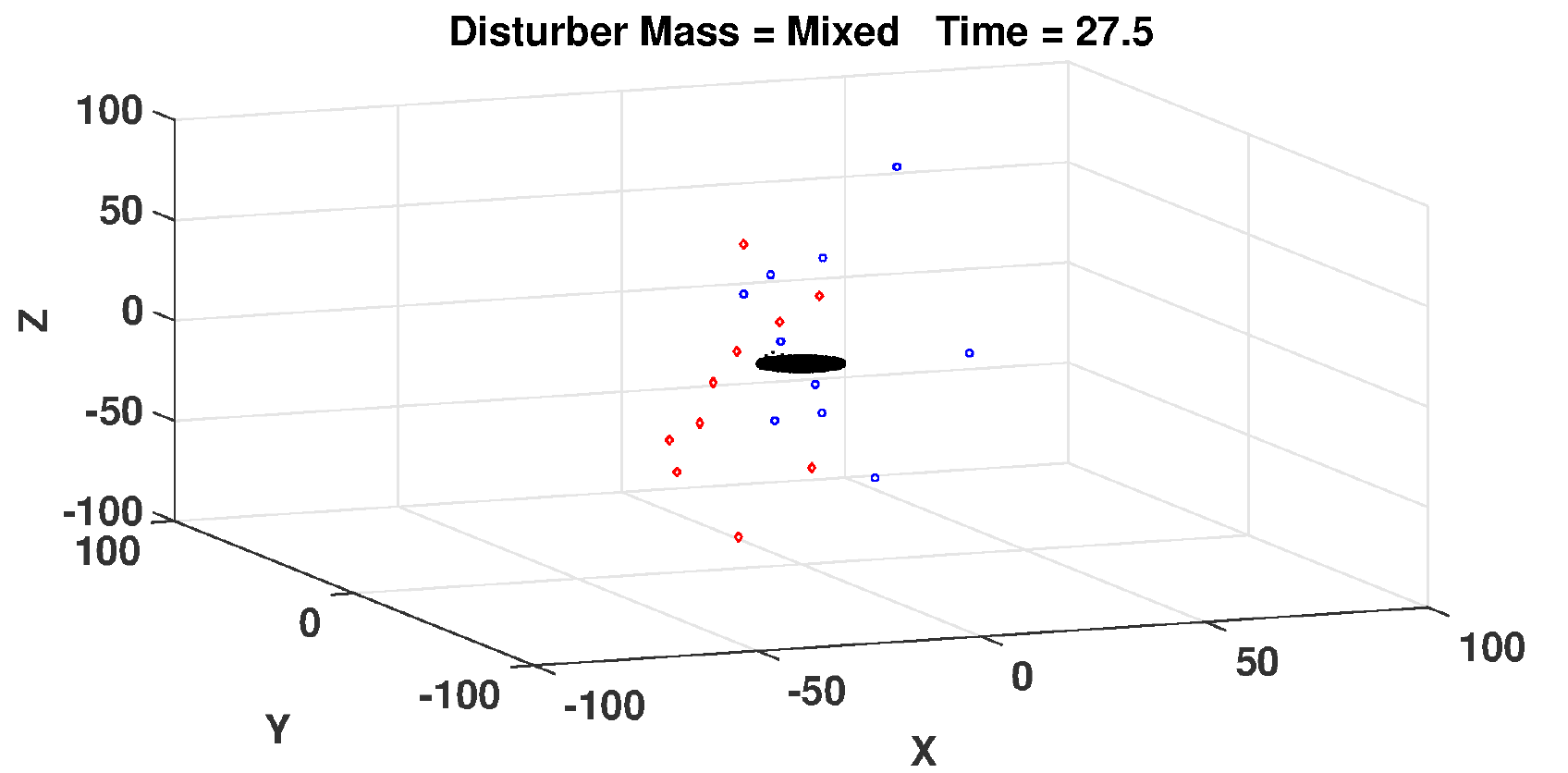}
\caption{Satellite alignment as defined by alignment criterion at 269 Myr for a run with a realistic mix of satellite masses. (Run \#23)}
\label{fig:figure16}
\end{figure}

\begin{figure}
\includegraphics[width=\columnwidth]{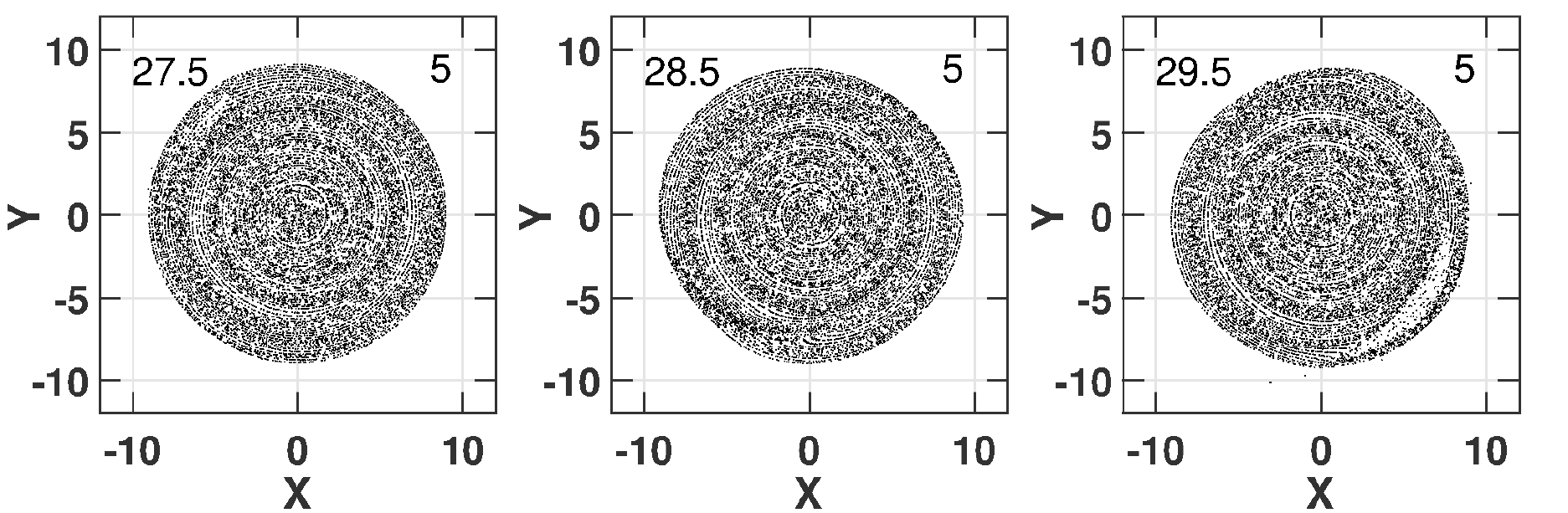}
\caption{Spiral wave formation occuring over 20 Myr as a result of satellite alignment in Figure~\ref{fig:figure16}.}
\label{fig:figure17}
\end{figure}

\noindent Here, $n$ is the number of aligned satellites, $m_i$ are the individual satellite masses, $\vec{r_i}$ are the individual satellite position vectors (which are projected onto the direction $\vec{R}$), $M$ is the large satellite mass, $\vec{R}$ is the position vector of the large satellite, and $G$ is the universal gravitational constant. 
Using this criterion, we implemented a function within our simulation code that, for each timestep, would iterate over 3D space with the criterion's primary positional vector and flag satellite groups that fulfilled the criterion. Figs.~\ref{fig:figure10} -~\ref{fig:figure15} show flagged satellite alignments and the resulting spiral wave formation after the alignment has occurred. Figs. ~\ref{fig:figure10} and ~\ref{fig:figure11} use satellite masses similar to that of the criterion's large satellite; these results illustrate the ability of similar-mass satellites at larger radii to drive spiral waves in the disc. Figs. ~\ref{fig:figure12} -~\ref{fig:figure15} use satellites similar to the small satellites found around Andromeda \citep{Collins2015}, and are demonstrative of the ability of observed band structures to fulfill this criterion. Finally, Figs. ~\ref{fig:figure16} -~\ref{fig:figure17} use a mix of satellite masses to better emulate observed band structures. These latter runs were performed at higher band radii to probe the effectiveness of the alignment criterion for more distant, less dense band structures. The spiral wave behavior driven by alignments in these larger structures is, while diminished in amplitude, still present in the disc. 



In other circumstances, such as a spiral provoked by a single close flyby of a large satellite, the induced wave would dissipate in short order - \citet{Widrow2015} showed that such perturbations quickly die out without continued stimulus. In this case, however, the corotational nature of the band structure allows the satellites to continue aligning on a quasi-periodic basis, providing the continued impulses required to maintain the spiral in the disc. However, the irregularity of these alignments in both space and time leads to multiple interfering waves and flocculent spirals. While the wave continues to be present, it often becomes warped or damped as another portion of the disc is excited by satellites aligned in an inopportune location. In preliminary work \citep{Criswell2018}, which featured fixed-orbit satellites, we saw spirals that tended to be grand-design in nature. With free satellites, however, the spirals are often somewhat lopsided, discontinuous, or multi-armed.\\

\begin{figure}
	\includegraphics[width=\columnwidth]{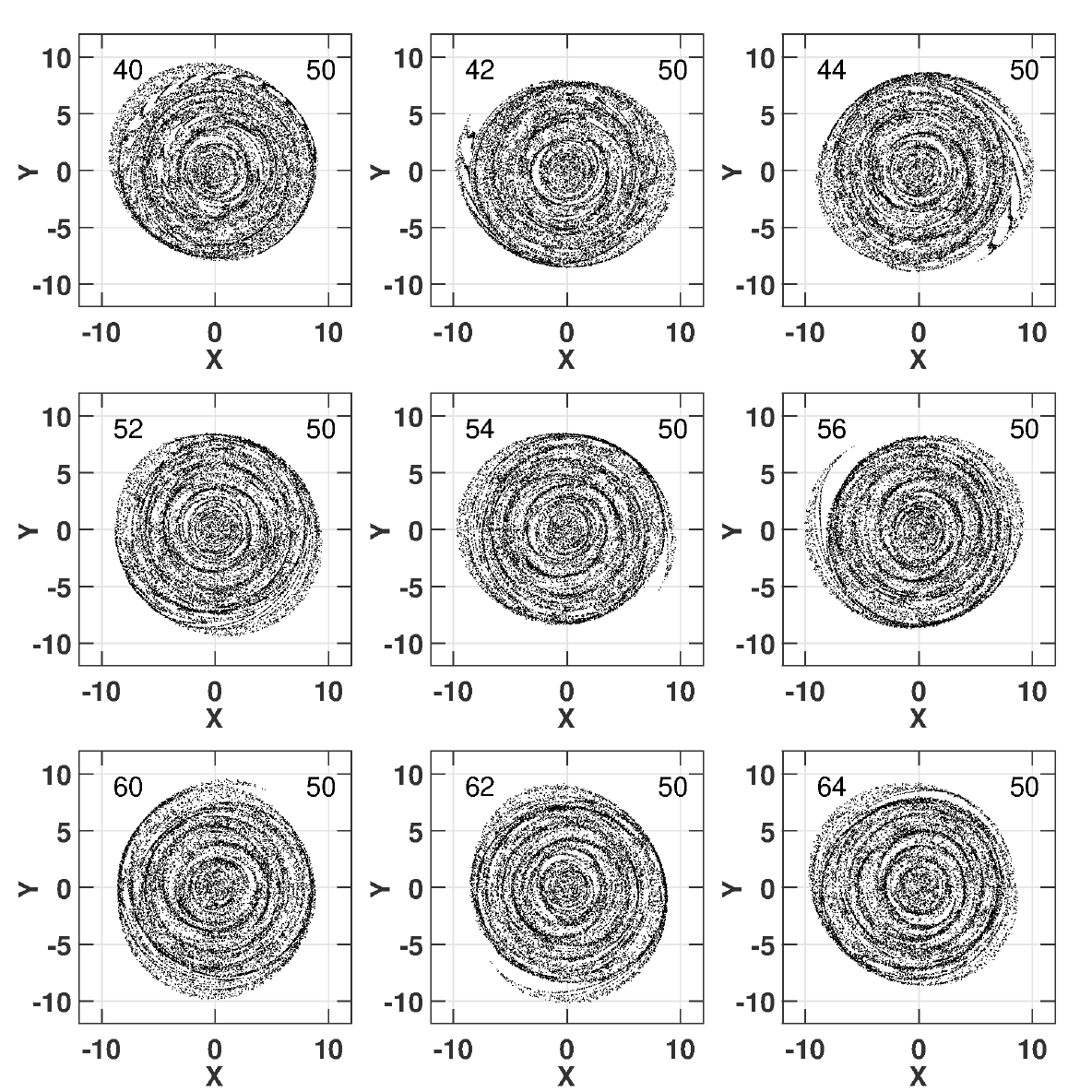}
    \caption{Face-on views of the disc, highlighting 3 moments of spiral wave re-excitation over 240 Myr in run \#14. Note that in each case, the spiral has decreased in intensity in the first frame and is reinvigorated over the course of the next two frames. Each re-excitation was preceded by or concurrent with an alignment identified by the criterion.}
    \label{fig:figure18}
\end{figure}

In order to judge whether satellite alignments could successfully and consistently predict the creation or reinvigoration of spirals, runs \#12-17 were performed at a higher time resolution \textemdash \ outputting frames for every 2 Myr of simulation time \textemdash \ allowing us a closer look at the moments immediately surrounding an alignment. The results were striking: new or reinvigorated spiral wave behavior in the disc was invariably connected with an alignment being present in the current step or the step prior, and the incidence of alignments again proved to be an excellent predictor of such behavior. Moreover, the increased time resolution allows us to view the hypothesized driving behavior. Figure~\ref{fig:figure18} shows examples of this, wherein a spiral wave flares up, decays, and is re-excited cyclicly over the course of 240 Myr.

\begin{figure}
\includegraphics[width=\columnwidth]{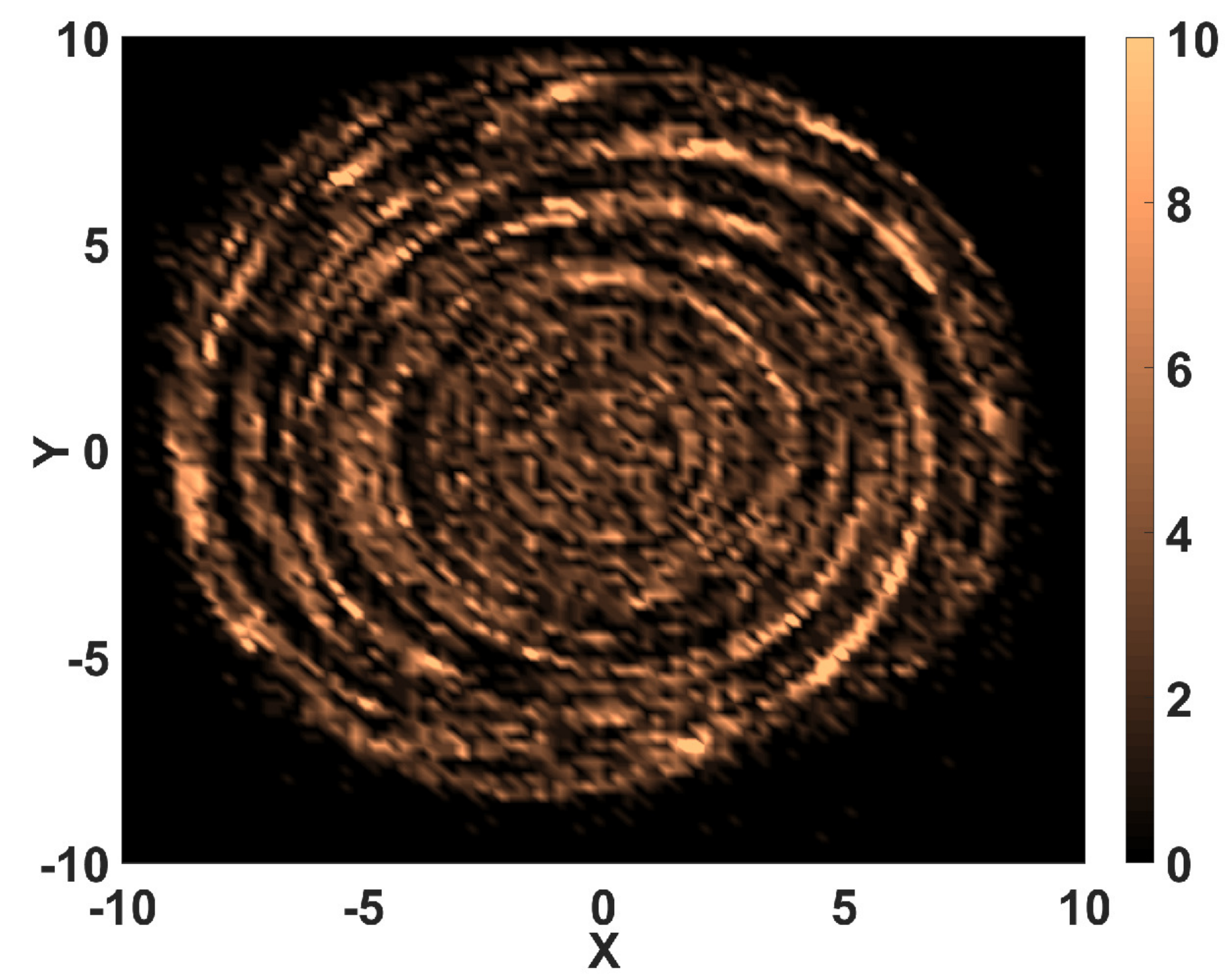}
\caption{Relative particle density of the disc at 598 Myr in run \#22. Overdense regions in the disc have ~3-4 times the density of the disc background.}
\label{fig:figure19}
\end{figure}

Additionally, for runs \#20-24, we plotted the particle density of the disc in order to ascertain an estimate of the relative density of the induced spiral waves compared to the disc background. Due to the abstraction of our `star' particles, the scale of these plots is solely intended as a relative measure rather than an absolute one. The spiral waves produced by band structure alignments tend to possess ~300-400\% the background density of the disc (see Fig. ~\ref{fig:figure19}). This exceeds observed values of ~$10-20\%$ for many-armed/flocculent spirals \citep{Thornley1996, Elmegreen2011}. However, this overestimate is expected; self-gravitation and disc damping in future work will reduce this value.

\subsection{Disc Thickening}
\begin{figure}
	\includegraphics[width=\columnwidth]{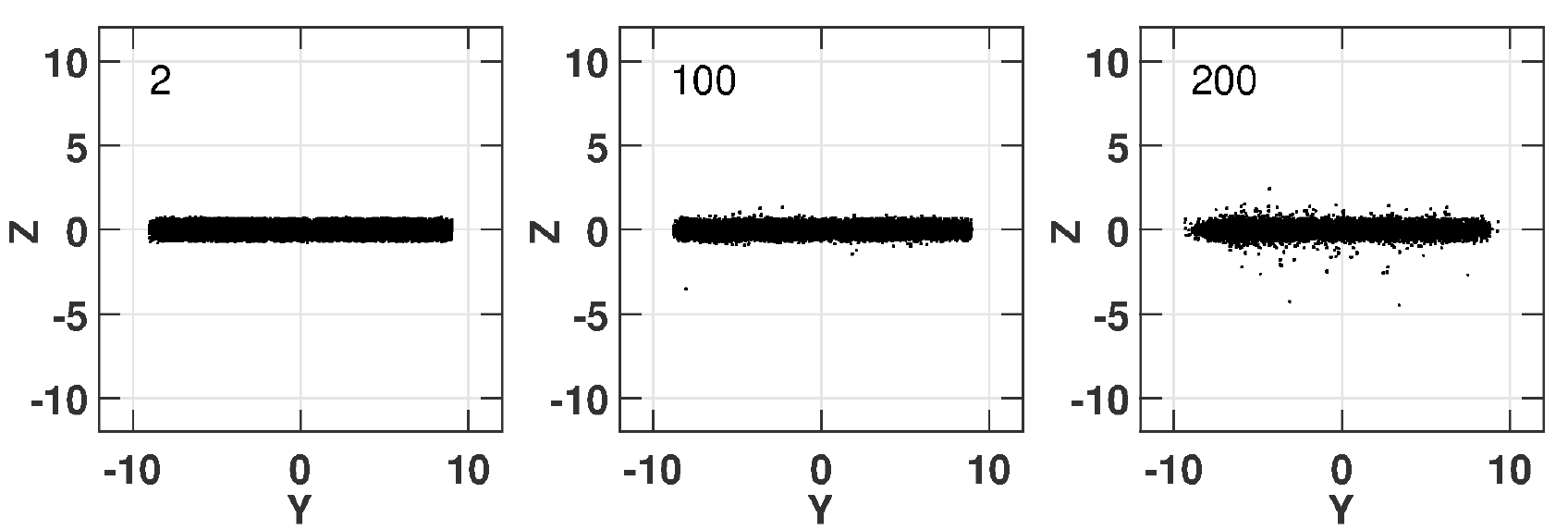}
    \caption{Side views of the disc at 20 Myr, 980 Myr, and 1.96 Gyr for a run with $1.4x10^9 M_{\sun}$ satellites. (Run \#16)}
    \label{fig:figure20}
\end{figure}

\begin{figure}
	\includegraphics[width=\columnwidth]{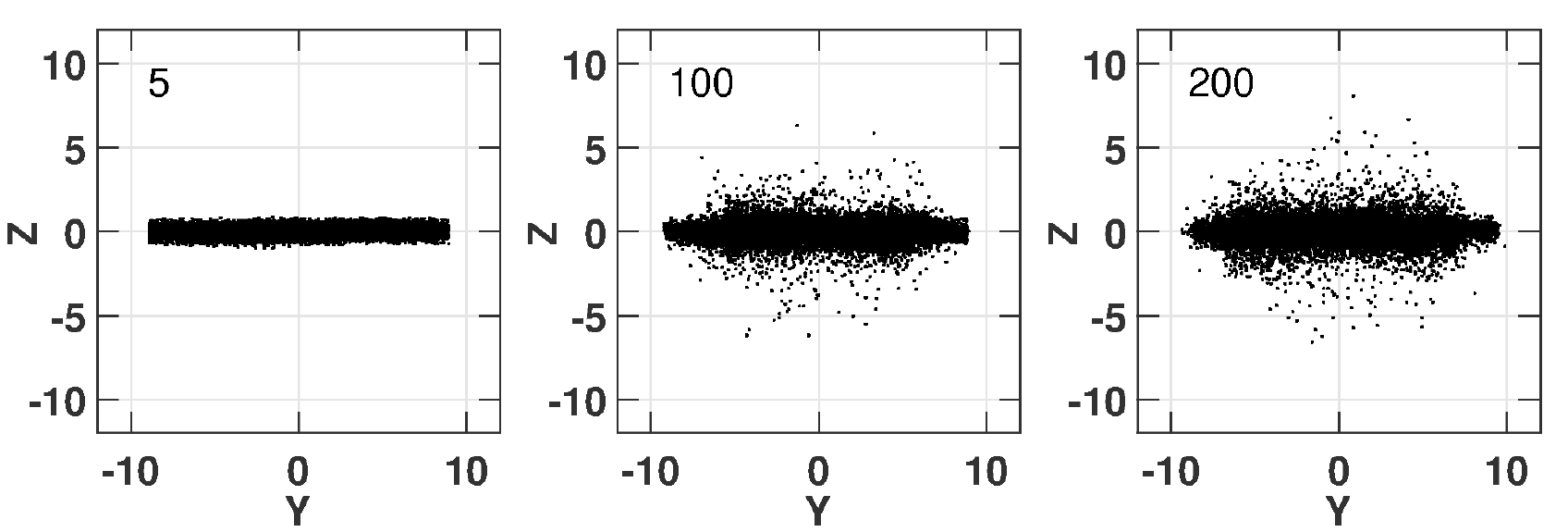}
    \caption{Side views of the disc at 49 Myr, 980 Myr, and 1.96 Gyr for a run with $6.9x10^9 M_{\sun}$ satellites. (Run \#6)}
    \label{fig:figure21}
\end{figure}

\begin{figure}
	\includegraphics[width=\columnwidth]{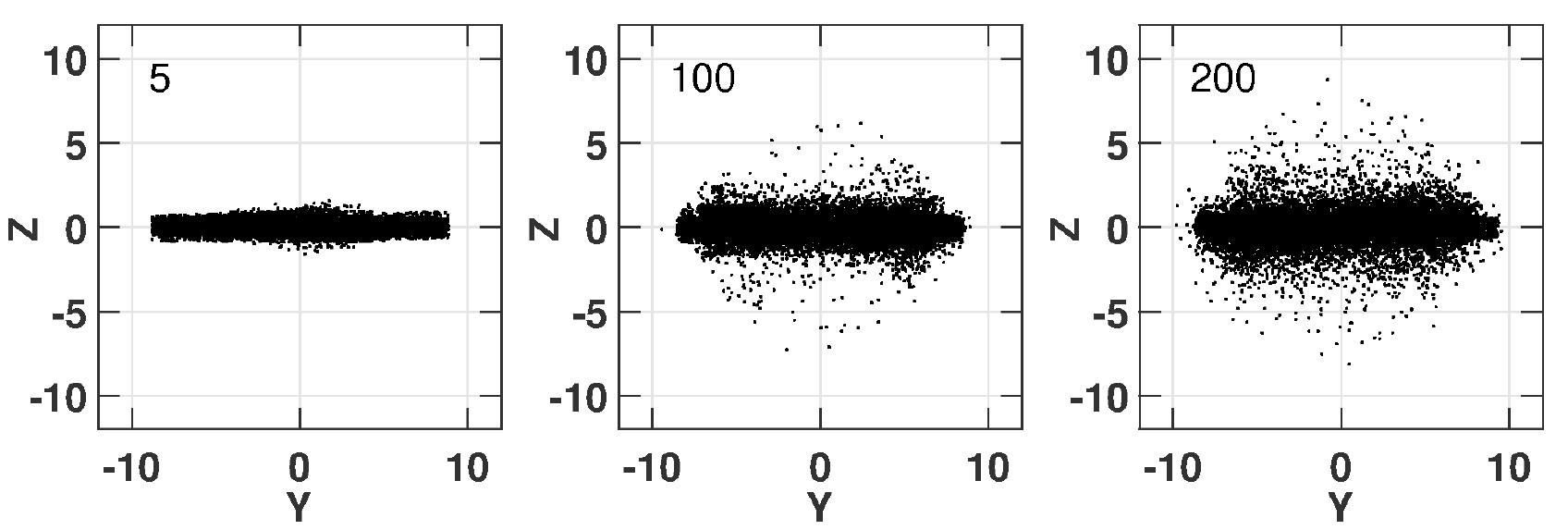}
    \caption{Side views of the disc at 49 Myr, 980 Myr, and 1.96 Gyr for a run with $13.8x10^9 M_{\sun}$ satellites. (Run \#11)}
    \label{fig:figure22}
\end{figure}

In the case of a number of different satellites, the multiple individual perturbations vary enough from particle to particle to create an effect roughly equivalent to scattering. This, as seen in previous work of \citet{Struck2017}, is a mechanism for inducing disc thickening and disc profile evolution. As in \citet{Criswell2018} for prescribed satellite orbits, we observed disc thickening in every simulation run performed, though the extent to which the disc thickens is highly mass-dependent. Figs.~\ref{fig:figure20},~\ref{fig:figure21}, and~\ref{fig:figure22} show the disc vertical evolution for satellites with masses ranging from $1.4x10^{9}M_{\sun}$ to $13.8x10^{9}M_{\sun}$. It is evident that large individual satellite masses on the scale of the LMC cause extreme vertical evolution, but satellites on a more reasonable scale for the majority of dwarf galaxies induce a lesser, yet still noticeable amount of thickening. Note, however, that the specific magnitude of the thickening depends on the fixed disc potential used in the code.


\subsection{Surface Density Profile Evolution}
\begin{figure}
	\includegraphics[width=\columnwidth]{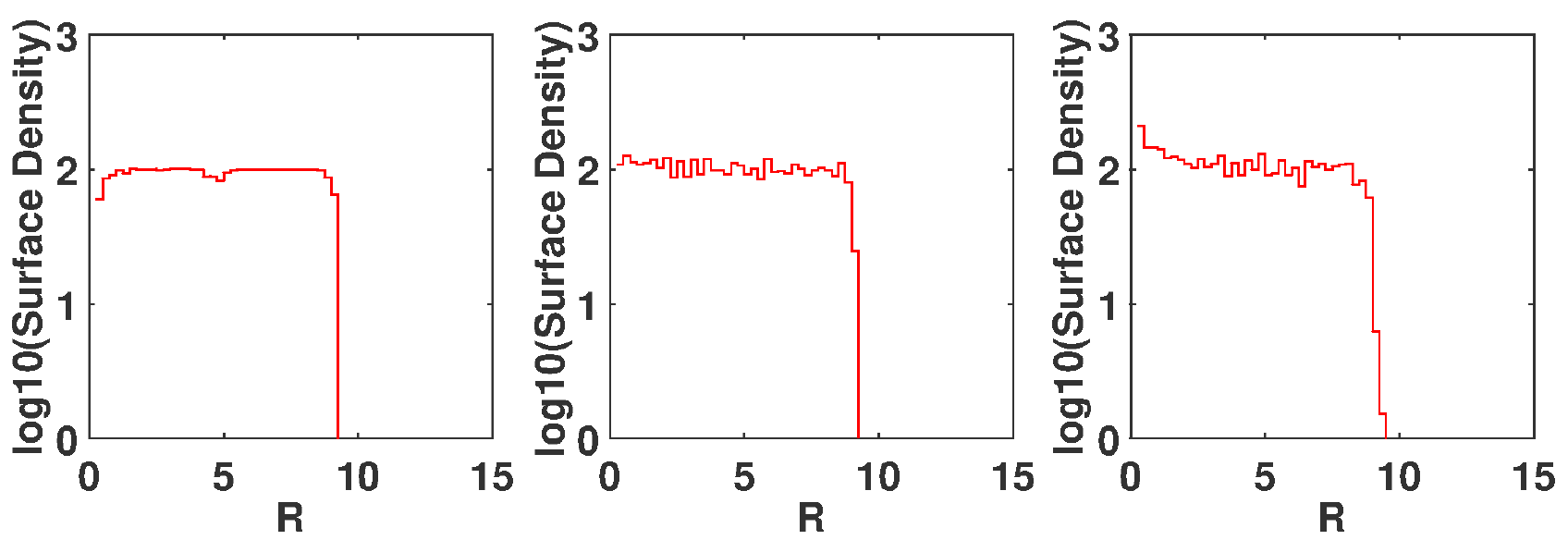}
    \caption{Disc density profile at 0 Gyr, 0.98 Gyr, and 1.96 Gyr for a run with $1.4x10^9 M_{\sun}$ satellites. (Run \#16)}
    \label{fig:figure23}
\end{figure}

\begin{figure}
	\includegraphics[width=\columnwidth]{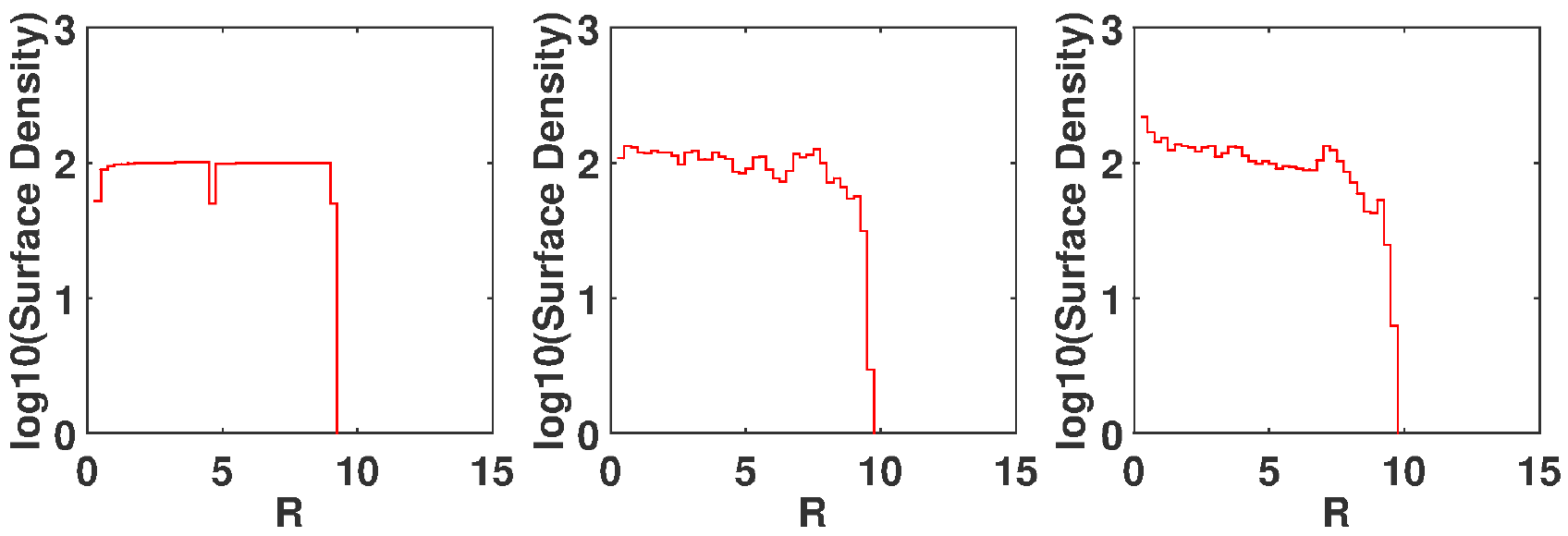}
    \caption{Disc density profile at 0 Gyr, 0.98 Gyr, and 1.96 Gyr for a run with $6.9x10^9 M_{\sun}$ satellites. (Run \#6)}
    \label{fig:figure24}
\end{figure}

\begin{figure}
	\includegraphics[width=\columnwidth]{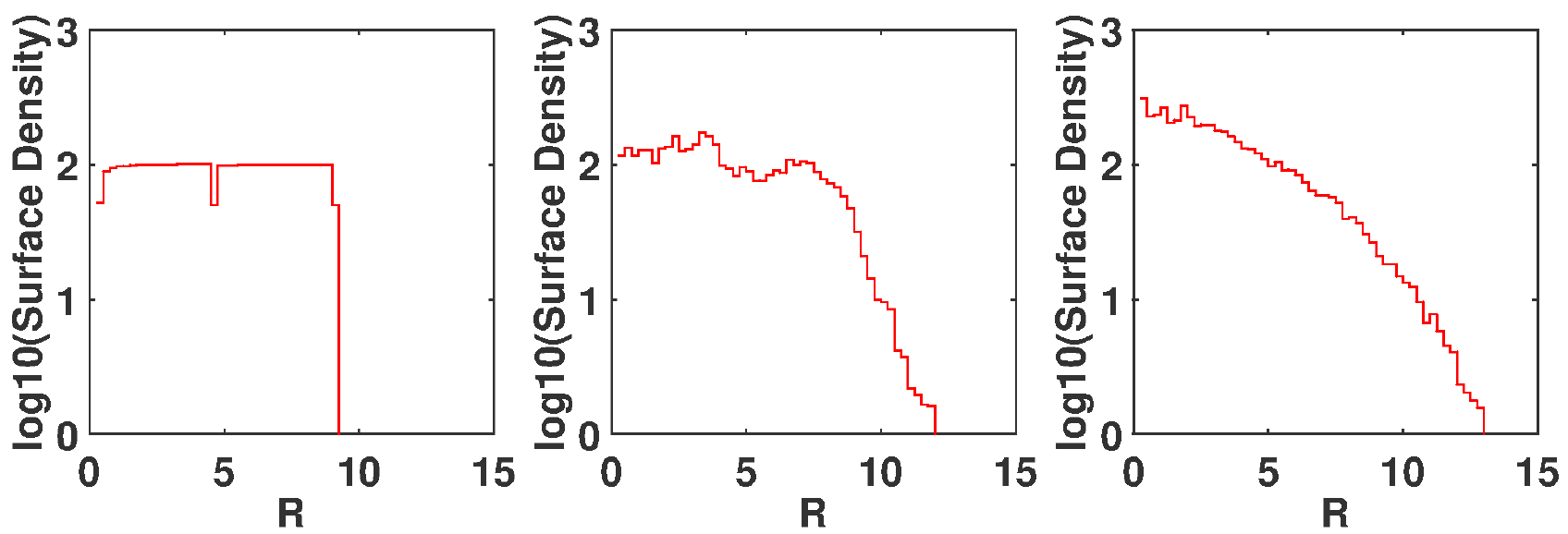}
    \caption{Disc density profile at 0 Gyr, 0.98 Gyr, and 1.96 Gyr for a run with $27.6x10^9 M_{\sun}$ satellites. (Run \#3)}
    \label{fig:figure25}
\end{figure}

One of the original purposes of this research, stemming from the Struck and Elmegreen work, was to see if band structures could promote formation of an exponential surface density profile within the disc. While the presence of a band structure does indeed contribute to the evolution of such profiles, the characteristic time frame for this contribution is long unless the individual satellite masses are quite large. This can be seen in Figs.~\ref{fig:figure23},~\ref{fig:figure24}, and~\ref{fig:figure25}, which show the disc profile evolution, from an initially flat profile, for satellites with masses ranging from $1.4x10^{9}\ M_{\sun}$ to $27.6x10^{9}\ M_{\sun}$. Due to the slow profile evolution driven by smaller masses in the typical dwarf galaxy range, band structures will not generally modify the surface density profile.


\section{Conclusions}

Groups of coplanar, corotating, low-mass dwarf companions can drive a number of phenomena within their host galaxy's disc. Stochastic alignments of these satellite galaxies are able to emulate the gravitational influence of a significantly more massive satellite and prompt the formation and repeated re-excitation of spiral waves within the disc. These alignments occur quasi-periodically, gathered and quickly dispersed by the varying orbital periods of their corotating component dwarfs. This quasi-periodicity results in flocculent, many armed spirals, as opposed to the grand design spirals seen when the alignments occur in more consistent intervals \citep{Criswell2018}. Moreover, alignments occur frequently enough that they are able to re-excite spirals on a significantly shorter timescale than the \citet{Widrow2015} damping effects by over a factor of 10 - alignments occured, on average, once every 20-30 Myr compared to that study's 200-800 Myr timescale (though generally with each alignment involving different satellites). Additionally, these alignments show an ability to drive stellar scattering, and therefore both vertical thickening and density profile evolution in the disc. While their contribution to the latter phenomena is small when compared to other factors, the band structure's influence on the disc's vertical evolution can be fairly extensive. 

While the effects of direct disc impacts by small satellites have been discussed at length \citep{Quinn1993,Dubinski2008, D'Onghia2016} and the ability of large companions to drive disc evolution is clear \citep{Sellwood1998,Bailin2004,Quillen2009}, we show that small companions in band structure alignments are able to drive spiral behavior and stellar scattering at a distance, without direct disc contact. Thus, our models provide evidence for the ability of low-mass, non-impacting companions to effect disc features and evolution over Gyr timescales. 

A potential concern for these results exists in the integrity of the individual dwarf galaxies. Tidal disruption of the band's component dwarfs by their host could limit the extent to which these small satellites are able to drive the behavior observed. As the simulation does not take this into account \textemdash the satellite dwarfs are not fully resolved \textemdash we collected 25 sets orbital data over the 2 Gyr simulation runtime for band structure satellites ranging in mass from $0.28$-$6.91 x10^{9} M_{\sun}$. In order to ascertain the rate of tidal disruption, we applied an approximate threshold force ratio, 

\begin{equation}
     $$\frac{F_H}{F_D} \approx 2\frac{M}{m}(\frac{R}{d})^3$$
	\label{eq:tidalthreshold}
\end{equation}

Here, $F_H$ is the tidal force from the central galaxy on a star in the outer reaches of a satellite dwarf. $F_D$ is the force exerted on the same star by the dwarf galaxy it occupies. $M$ and $m$ are the mass of the central galaxy and the dwarf, respectively; $R$ is the radius of the dwarf galaxy (e.g., \textasciitilde3 kpc) and $d$ is the distance between the satellite and the primary. If this ratio is $\geq$ 1, then the dwarf satellite is considered to have experienced a disruption event. 

We found that, for the band structure radii used in this study, the satellite dwarfs needed to be at least $2$-$3 x10^9 M_{\sun}$ (or very compact) to avoid the majority of the structure being tidally disrupted. An average of 69\% of $2.8x10^8 M_{\sun}$ dwarfs were repeatedly disrupted, compared to 55\% for $1.4x10^9 M_{\sun}$ dwarfs, 36\% for $2.8x10^9 M_{\sun}$ dwarfs, and 23\% for $6.9x10^9 M_{\sun}$ dwarfs. These disruption events occured, on average, once every 74 Myr for $2.8x10^9 M_{\sun}$ dwarfs and once every 162 Myr for $6.9x10^9 M_{\sun}$ dwarfs. However, in these higher-mass cases, the remaining dwarfs only experienced disruption events at a average rate of \textasciitilde1 per 4 or 5 Gyr, respectively. While these results do set a lower limit on the dwarf galaxy masses that can drive the effects seen in this study over long timescales, the majority of masses observed in these structures likely fall above the threshold for significant disruption \citep{Collins2015}.

The results of this study further inform our understanding of how galaxies can be influenced by their local environment and also pose an interesting observational question: given that we have observed these structures around the Milky Way, Andromeda, and Centaurus A \citep{Kroupa2005,Koch2006,Mueller2018}, can we find satellite alignments - current or past - and attempt to connect their influence with observed phenomena within their host galaxy's disc? A number of additional avenues for theoretical research remain, especially concerning the long-term evolution of the satellite band. While this work addresses in part the issue of tidal disruption, a thorough investigation of its effects would require full N-body simulations that resolve the satellite dwarfs. Other effects, such as dynamic friction and gravitational interactions between individual satellites and characterization of the spiral wave behavior in the disc are also of interest. The quasi-periodic nature of the satellite alignments in this study relies on the band structure's component bodies remaining approximately corotational and coplanar. Full N-body simulations that resolve both the central disc and its satellites are needed to address these issues fully. 

\section*{Acknowledgements}

The authors would like to acknowledge use of NASA's Astrophysics Data System ADS, as well as many helpful conversations with T. Yaeger and J. Wu. 







%
%


\bsp	
\label{lastpage}
\end{document}